# Cellular blood flow modelling with HemoCell

*Gabor Zavodszky[1*], Christian Spieker[1], Benjamin Czaja[2], and Britt van Rooij[3]*

[1]University of Amsterdam, Science Park 900, 1098XH, Amsterdam, Netherlands;

[2]SURF Cooperation, Science Park 140, 1098 XG Amsterdam, The Netherlands;

[3]Philips Medical Systems, Veenpluis 6, 5684 PC Best, The Netherlands;

Corresponding Author
*E-mail: g.zavodszky@uva.nl

**ABSTRACT**

Many of the intriguing properties of blood originate from its cellular nature. Bulk effects, such as viscosity, depend on the local shear rates and on the size of the vessels. While empirical descriptions of bulk rheology are available for decades, their validity is limited to the experimental conditions they were observed under. These are typically artificial scenarios (e.g., perfectly straight glass tube or in pure shear with no gradients). Such conditions make experimental measurements simpler, however, they do not exist in real systems (i.e., in a real human circulatory system). Therefore, as we strive to increase our understanding on the cardiovascular system and improve the accuracy of our computational predictions, we need to incorporate a more comprehensive description of the cellular nature of blood. This, however, presents several computational challenges that can only be addressed by high performance computing. In this chapter we describe HemoCell[1], an open-source high performance cellular blood flow simulation, which implements validated mechanical models for red blood cells and is capable of reproducing the emergent transport characteristics of such a complex cellular system. We discuss the accuracy, the range of validity, and demonstrate applications on a series of human diseases.

---

[1] https://www.hemocell.eu

# 1  The cellular properties of blood

Blood is strongly linked to many of the physiological processes in the human body. Its major three functions can be categorised as transportation, protection, and regulation [1]. It transports oxygen, nutrients, and various wastes of the metabolic functions to maintain normal functioning of tissues all around our body. It also takes a significant role in most of our protective functions, including immune processes and haemostatic mechanisms. Finally, it regulates the balance of body fluids, cellular pH tension, and maintains overall thermobalance.

To be capable of providing its many functions, the composition of blood is far from simple. It is a dense suspension of various cells immersed in blood plasma, such as red blood cells (RBCs), platelets (PLTs), and white blood cells (WBCs) [2]. Blood plasma is usually regarded as a Newtonian incompressible fluid containing water and a series of proteins, hormones, nutrients, and gases. Under some circumstances the effect of these solutes can become significant and lead to non-Newtonian effects such as plasma hardening [3]. In the current chapter we will not consider these plasma effects. RBCs are by far the most numerous of the cells in blood with a normal volume fraction or haematocrit of 45% in adults. They have a biconcave shape that has a large surface to volume ratio, which facilitates the exchange of oxygen by increasing the reaction surface. The diameter of these cells when undeformed is 6-8 μm with a 2-3 μm thickness. Their deformability originates from their structure that consist of a lipid bilayer that can dynamically attach and detach from the underlying supporting cytoskeleton formed by elastic spectrin proteins connected to each other via actin filaments. At low strain the membrane behaves similarly to elastic solids, however, at high shear deformation it behaves more like a fluid [4]. The intracellular fluid, the cytosol, is composed of a haemoglobin solution that has a five times higher viscosity compared to blood plasma. A single drop of blood contains approximately 150 million RBCs (see Fig. 1).

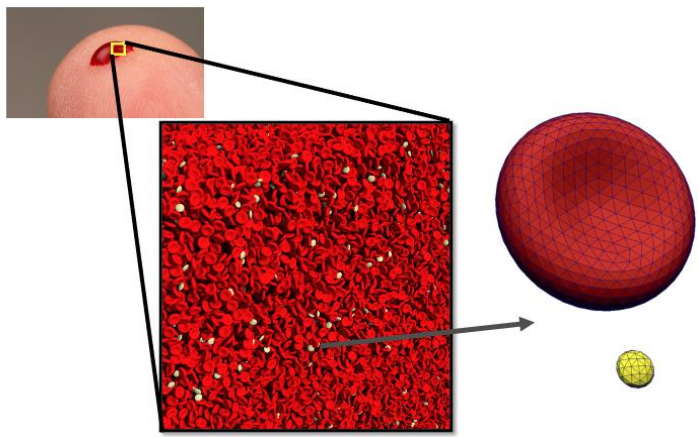

*Figure 1* A drop of blood modelled as a suspension of red blood cells and platelets.

The dynamics and deformability of RBCs is important, as they give rise to the unique bulk properties of blood. The general behaviour of these cells in (homogeneous) shear can be characterized with three separate regimes. At low shear rates RBCs tumble and flop retaining their original biconcave shape [4]. The increase of shear leads to a swinging motion regime, and finally at higher shear rates the membrane deforms, (partially) detaches from the cytoskeletal structure, and starts to rotate around like the tread of a tank [5]. In such higher shear flows the RBCs tend to line up and align with the flow, which in turn leads to a drop in bulk viscosity. This shear-thinning behaviour was observed in pure shear (far from walls) in the well-known Chien experiments [6]. These also showed a steep increase in viscosity as the shear rate falls below 10 $s^{-1}$, that later turned out to be caused by the aggregation of RBCs into column-like structures (rouleaux) [7], which has been shown to be induced by external plasma protein pressure [8]. The tendency of RBCs to migrate away from the vessel walls [9] leads to a different variety of shear-thinning. As blood flows through smaller vessels, the apparent viscosity drops, as denoted by the Fåhraeus-Lindqvist effect. As the RBCs vacate the vicinity of the vessel wall, they create a plasma rich, red cell free layer (CFL), which can effectively act as a lubrication layer. As we go towards smaller vessels the relative effect of this layer grow, leading to an overall decreasing viscosity [10]. These observations were followed by decades of experiments that were later aggregated and shown to be consistent in the famous work of Pries [11]. As RBCs move towards the center of the vessel, they create a highly non-homogeneous haematocrit distribution that peaks in the middle of the channel and goes to zero next to the wall (CFL). This results in an uneven viscosity distribution and causes a departure from the Poiseuille-profile that characterizes stationary Newtonian channel flows [12].

Given that these processes were studied mainly under synthetic conditions (in straight glass tubes, or in pure shear) that are not representative of the human circulation, the full extent of these phenomena are still not completely known. The suspension nature of whole blood therefore needs to be taken into consideration in order to properly study the processes, both physiological and rheological, that occur in blood flow.

## 2   Methods - accurate computational modelling of blood flows

### 2.1   Simulating blood on a cellular scale

Resolving the complete rheology of blood flows in experimental (*in vitro* or *in vivo*) settings has several limitations. These are primarily caused by the limited spatiotemporal resolution of our current measurement technology, and by the sheer amount of information necessary to capture the full characteristics of cellular blood. One way to overcome these limitations is by applying cellular flow simulations, that is, moving the experiments *in silico*. In recent years many numerical approaches have been developed to simulate the cellular nature of blood [13]. These typically follow the same structure and contain three major components: a fluid model that reproduces the flow of plasma, a mechanical model that describes the deformation of cells, and fluid-structure interaction method that couples these two efficiently. The choice of numerical technique and its implementation varies, and most combinations have specific advantages and disadvantages that can make them suitable for a given research question. Some of the most widely used solutions include discrete particle dynamics (DPD) [14], lattice Boltzmann method (LBM) in combination with finite element (FEM) cells coupled by the immersed boundary method (IBM) [15], or smoothed particle hydrodynamics (SPH) for all three components [16]. In the following we describe HemoCell, an open-source cellular blood flow simulation code that is designed with the aim of reproducing various microfluidic scenarios to complement experimental measurements. In order to reach this goal, the choice of numerical methods should fulfil a set of requirements:

1. *Enable large-scale flows.* In order to match experimental settings including the geometry (e.g., in a bleeding chip), a large number of cells must be simulated efficiently. Therefore, HemoCell was built with great scalability in mind to allow efficient simulation of a few dozen cells on a laptop up to millions of cells on the largest supercomputers.
2. *Stability at high shear rates.* Many of the investigated phenomena include biomechanical processes as a fundamental component (e.g. initial stages of thrombus formation). To capture these, the simulations need to reproduce realistic flow velocities that often translate to high shear rates locally. A unique feature of HemoCell is that the computational methods are fine-tuned to allow numerical stability at sustained high deformations.
3. *Advanced boundary conditions.* Matching microfluidic scenarios necessitate the support of a series of boundary conditions that can generate constant cell influx, produce pure shear, or enable rotating boundaries in a cellular suspension. Furthermore, the solution needs to be able to represent complex geometric boundaries, such as the surface of a growing thrombus.

To fulfil these requirements, HemoCell is designed with three main components: lattice Boltzmann method for the plasma flow, discrete element method (DEM) for the cellular mechanics, and IBM to create a flexible and accurate coupling between these. The structure of HemoCell is outlined in Fig. 2.

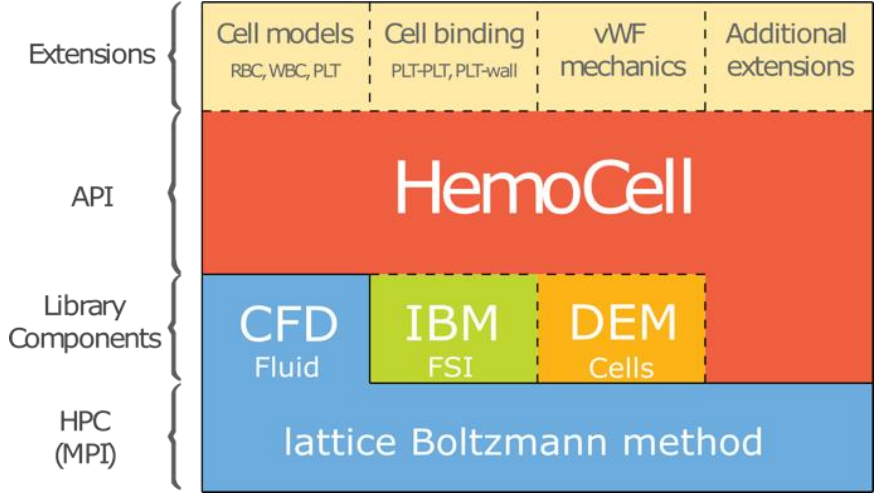

*Figure 2 Outline and structure of the major components of HemoCell.*

**2.2** Simulating fluid flow with the lattice Boltzmann method

Blood plasma in HemoCell is modelled as an incompressible Newtonian fluid, and the governing equations (Navier-Stokes) are solved with the lattice Boltzmann method. This method is known to be able to accurately capture flow in complex vascular geometries and it is well suited to high-performance parallel execution [17].
Historically, this method is an incremental evolution of the lattice gas automata, and the theoretical bases were formulated by Bhatnagar, Gross and Krook (BGK) [18]. For a detailed review the reader is referred to the work of Chen [19]. Using their collision operator, the core equation of the LBM method represents distribution functions in lattice space (i.e., discrete time, space, and velocity directions):

$$f_i(\vec{x} + \vec{e_i}, t + 1) = f_i(\vec{x}, t) + \frac{1}{\tau}\left(f_i^{eq}(\vec{x}, t) - f_i(\vec{x}, t)\right),$$

where the $i$ running index denotes the possible discrete velocity directions given by the actual grid model, $f_i^{eq}$ represents the equilibrium distribution function, $e_i$ is the direction of the selected velocity and $\tau$ is the relaxation time of the kinetic system. The fluid density $\rho$ and the macroscopic velocity $\vec{u}$ can be recovered at any lattice site from the first two moments of the distribution function:

$$\rho = \sum_i f_i$$

$$\vec{u} = \frac{1}{\rho}\sum_i f_i \vec{e_i}$$

The equilibrium distribution function ($f_i^{eq}$) follows the Boltzmann distribution in the form described by [20]:

$$f_i^{eq}(\vec{x}, t) = w_i \rho \left[1 + 3(\vec{e_i}\vec{u}) + \frac{9}{2}(\vec{e_i}\vec{u})^2 - \frac{3}{2}\vec{u}^2\right],$$

where $w_i$ denotes the numerical grid dependent weight values. Applying the expansion described by Chapman and Enskog in the limit of long wavelengths (or low frequencies) [21], from the above-defined system the Navier-Stokes equation for incompressible flows can be recovered with an ideal equation of state: $p(\rho) = \rho c_s^2$ and a kinematic viscosity of $\nu = c_s^2 \left(\tau - \frac{1}{2}\right)$ where $p(\rho)$ stands for the pressure while $c_s$ is the numerical grid-dependent speed of sound which takes the value of $\frac{1}{\sqrt{3}}$ for the most often used grids (e.g., D3Q19). For an in-depth description of the lattice Boltzmann method, including its recent advancements the reader is referred to the book of [22].

### 2.3 The computational model of the cells using the immersed boundary method

The fluid component, described by LBM, operates on an Eulerian grid, while the cells are represented by a triangulated membrane mesh (see the cells in Fig. 1), whose surface vertices have a Lagrangian description. The two methods are coupled together by the IBM developed by Peskin [23]. This is an explicit coupling through external force terms: the cells deform due to the motion of the fluid field, and these deformations yield a non-stress free cell state described by constitutive models. The force response to these deformations is applied to the flow as an external force, influencing the dynamics of the flow. Since the fluid and cellular components are described on different numerical grids, the coupling steps include linear interpolation in both of these steps. A crucial question is how to describe the force response of cells to deformation. There are several existing solutions [24], [25] all based on approximating the mechanics of the cytoskeletal structure of the cells. In HemoCell this fundamental idea is expanded to differentiate between small deformations that are expected to yield linear reaction based on the response of the bilipid membrane, and large deformations that yield highly non-linear response that combines contribution from both the membrane and the cytoskeletal structure. The overall force response of this constitutive model is comprised of four major components:

$$F_{total} = F_{link} + F_{bend} + F_{area} + F_{volume}.$$

These force components correspond to a divers set of possible deformations of the cell and its membrane. The first component ($F_{link}$) relates to the stretching of the spectrin filaments in the cytoskeletal structure. The second one ($F_{bend}$) accounts for the bending rigidity of the membrane combined with cytoskeletal response for large deformations. The third ($F_{area}$) maintains the incompressibility of the bilipid membrane, while the last one ($F_{volume}$) ensures quasi-incompressible volume. The parameters of these forces are derived from limiting behaviour of the cells and fitted to experimental measurements (including optical tweezer stretching, Wheeler test). Finally, the free parameters are finetuned for numerical stability. A more in-depth description of these forces and their numerical implementation can be found in [26]. The resulting computational model was thoroughly validated in a series of single-cell and many-cell experiments for both healthy and diabetic blood [27]. To ensure the robustness of the model it was subjected to detailed sensitivity analysis and uncertainty quantification [28].

### 2.4 Creating initial conditions for cellular flow

Creating the initial conditions for a cellular suspension is a challenging task. Under physiologic conditions stationary blood is well-mixed, the position and the orientation of the cells are both random uniform (see Fig. 3).

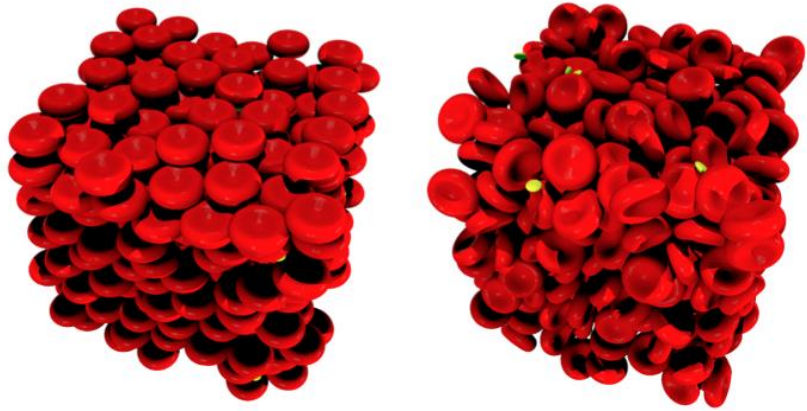

*Figure 3 Two randomized initial conditions. Left – only the positions are randomized. Right – both the positions and the orientations are randomized.*

Naïve randomization routines are only efficient in low-haematocrit regimes. As the volume ratio of cells increases, avoiding overlaps is becoming more difficult. This practically leads to a dense packing problem of irregular cellular shapes. Since HemoCell is designed with large-scale applications in focus, the solution to this problem must also be performant enough to be able to handle up to $O(10^6)$ cells of various types. The way to reduce computational cost efficiently, is to utilise surrogate objects that are easier to handle. In this case the implementation carries out a dense packing of ellipsoidal shapes, that each encompass a single cell. These ellipsoids are initialized with random position and orientation, after which an iterative method (the *force-bias method*) computes their overlap and applies repulsive forces between each overlapping pair proportional to their overlapping volume:

$$F_{ij} = \delta_{ij} p_{ij} \frac{\vec{r_j} - \vec{r_i}}{|\vec{r_j} - \vec{r_i}|},$$

where $\delta_{ij}$ equals 1 if there is an overlap between particle $i$ and $j$ and 0 otherwise, while $p_{ij}$ is a potential function proportional to the overlapping volume. This method is capable of efficiently initializing large, well-mixed cellular domains, while reaching high packing density (i.e., haematocrit) values [29].

## 2.5 Advanced boundary conditions

Every computational model that has a finite domain requires boundary conditions. These define the behaviour at the borders of the domain or at borders of different regimes within the domain. A trivial example implemented even in simple fluid flow simulations is the no-slip boundary condition. In the case of cell-resolved flow models like HemoCell, additional boundary conditions need to be introduced to cover cell-to-cell, cell-to-fluid and cell-to-domain boundary interactions. To enable diverse applications for HemoCell simulations, e.g. in more arbitrary geometries such as curved and bifurcated blood vessels, more advanced boundary conditions are necessary. Here, three boundary conditions implemented in the HemoCell framework, namely the periodic pre-inlet, the rotating velocity field boundary and the Lees-Edwards boundary condition [30] are discussed along with the more standard periodic boundary condition.

A periodic boundary condition is commonly used for cellular flow simulations in domains with high symmetry, such as a straight channel section with a single inlet and outlet (see e.g. [26]). By mimicking an indefinitely long tube or channel, computational costs can be saved without sacrificing spatial resolution, e.g. through downscaling. For an example application, a periodic boundary condition is applied in two dimensions to mimic the flow profile and cellular interactions at the center of a parallel plate flow chamber (see Fig. 4 A).

In many cases a periodic boundary condition cannot be implemented, for example in case of multiple outlets with only a single inlet, as is the case for a bifurcated vessel section. In situations where periodicity cannot be applied, such as in the case of a curved (or any arbitrarily shaped) vessel section, additional description is required for the in- and outflowing cellular distributions. To allow for a constant influx of cells, a periodic pre-inlet is attached in a serial fashion to the inlet of the domain of interest. This periodic pre-inlet is a straight channel with periodic boundaries in the direction of the flow [31]. It can be regarded as a small additional simulation prepending the domain of interest. Cells and fluid propagating across the outlet of the periodic pre-inlet (that is joined to the inlet of the main simulated domain), are duplicated into the main domain. Fig. 4 C visualizes this process in case of a curved vessel simulation by depicting the initial time-step of the empty main domain and filled pre-inlet and a later stage when the entire domain is filled with cells from the influx.

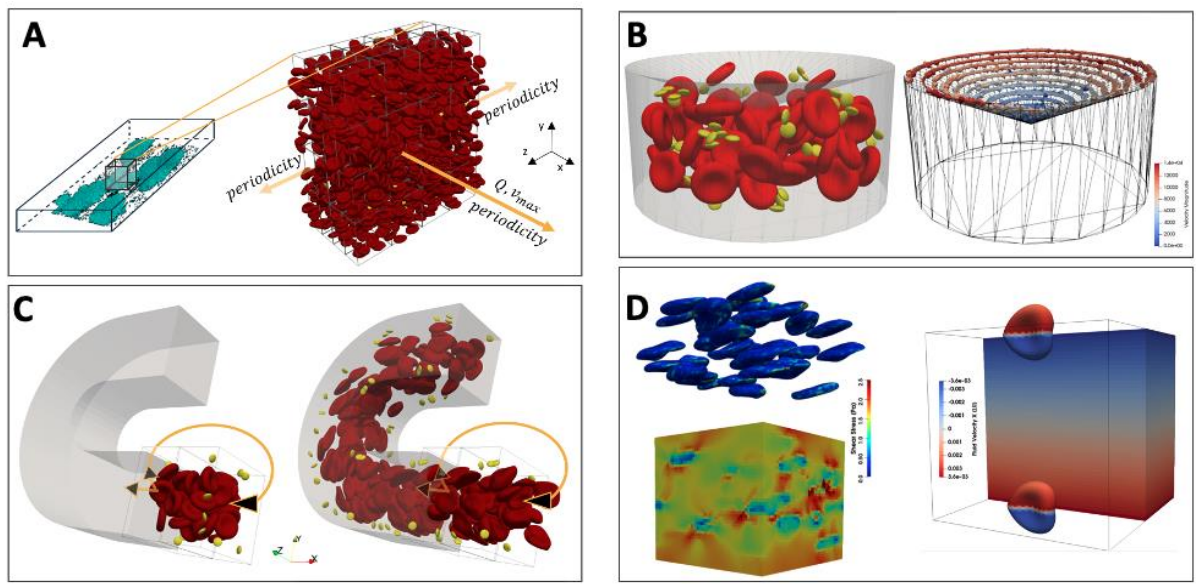

*Figure 4 Boundary conditions in HemoCell. **A)** periodic boundary condition. **B)** rotating boundary condition – in this case the cone-shaped top lid. **C)** Pre-inlet section to provide a continuous influx of well mixed cells. **D)** Lees-Edwards boundary condition to provide pure shear environment.*

Many hematological devices include moving or rotating components. One such example is a cylindrical device aiming to create constant shear in a volume to observe the shear induced behaviour of blood. This device allows the analysis of platelet deposition at the device's bottom plate due to a constant shear gradient which is maintained by a rotating cone top lid [32]. A virtual replica of this device can be created in HemoCell, which allows access to more detailed information on the process (see Fig. 4 B), such as the trajectory of the cells, or local fluid stresses. The top lid of the geometry is a cone shape indented at a desired angle between 0 - 45°. The rotation of this lid is implemented via a rotating velocity field at the surface of the cone shape. This results in a linear velocity profile at the cone surface which propagates through the cylinder and thus causes rotation of the fluid including the immersed cells.

Finally, the implementation of a uniform shear environment, such as in a Couette flow viscometer, allows for the calculation of bulk viscosity of blood and the diffusivities of the cells. This setup comes with the caveat that the presence of boundaries (wall or velocity boundary) induces lift force on the immersed cells, which in return causes the formation of a CFL [26]. One solution to this problem is to oversize the domain until the effects of the boundaries become negligible. In order to prevent this computationally expensive upscaling, Lees-Edwards boundary conditions are implemented to create a boundary-less uniform shear environment. The simulation is periodic in all dimensions, while imposing constant opposing velocities on two parallel boundaries in the direction of the shear. Figure 4 D (left) shows the stress patterns on the immersed cells and the embedding fluid, which are used to calculate the bulk viscosity. Figure 4 D (right) shows a single red blood cell crossing the Lees-Edwards boundary and hence being copied onto the other side of the domain, while experiencing forces from two opposing directions.

The advanced boundary conditions implemented in HemoCell widen the range of potential applications. The 3D periodic pre-inlet, the rotating velocity field boundary and the Lees-Edwards boundary condition are novel implementations for cellular simulations, which enable their application in complex geometries mimicking rheological devices and realistic vessels.

### 2.6 Performance and load-balancing

The largest scale deployment of HemoCell at the time of writing demonstrated the execution of a single simulation on over 330,000 CPU cores, while maintaining over 80% weak-scaling efficiency. Given the tightly coupled nature of the code involving both structured (LBM) and unstructured (DEM) grids this performance scaling can be regarded as excellent. To achieve this, several advanced techniques are applied.

The computation of the constitutive model of the cells is very costly (at the typical resolution a single RBC is simulated via a system of 5000 equations). Since the flow field and the cellular components utilise different numerical methods, their stability w.r.t. the time-step size also differs. The equations of cells are integrated by either the Euler or Adams-Bashforth method depending on the choice of the user. Both of these allow larger integration steps compared to LBM. For this reason, the computation of the cells is separated in time from that of the fluid field, allowing the constitutive model to be evaluated less frequently and save significant computational cost. The separation of integration time scales can be set to a constant value or can be set adaptively during the simulation depending on the scale of stresses in the system [29].

Another challenge that arises at scale is the change in computational load distribution. Cells move around relative to the stationary geometry. This means that they can vacate certain region or aggregate in others. The initial domain decomposition is based on the starting homogeneous state of the simulated domain, however, during the simulation this can increasingly become non-homogeneous. If this happens the original domain decomposition is no longer sufficient, since regions with no or very few cells compute much faster than regions with high cell density. This means that CPUs with less work will idle, reducing the parallel efficiency significantly. To circumvent this problem HemoCell has load-balancing capability, that monitor the fractional load-imbalance of the simulation, and when it surpasses a given threshold the simulation is paused (checkpointing), redistributed, and continued with the new domain decomposition [33]. This load-balancing step necessarily incurs additional computational cost, therefore the choice of the imbalance threshold is important.

# Notes - Applications of HemoCell

## 3.1 Cellular trafficking and margination

A major reason to apply microscopic blood simulation is to gain access to high-resolution information that is not possible through experimental techniques. For example, to obtain the detailed trajectories of cells, or correlate their positions with their deformation in order to explain emergent properties such as the formation of a cell free layer. Another important reason is to investigate a flow system in the presence of gradients. For instance, even a simple flow in a straight vessel segment (e.g., see Fig. 5) has various gradients that cannot be tracked accurately in an experimental setting [34]. These include gradients in the cell distribution, gradients in velocity and gradients in shear-rate (note: this flow is not Newtonian).

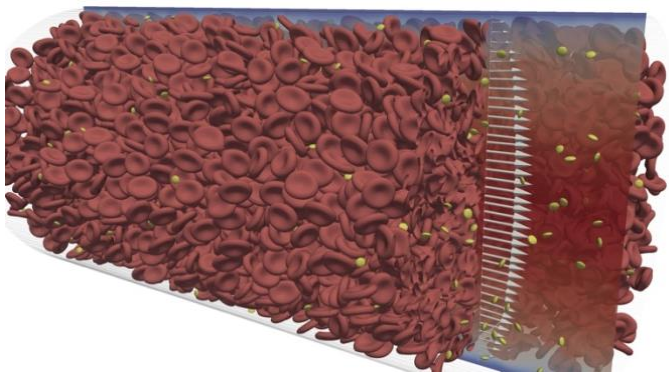

*Figure 5 Outline of cellular flow in a straight pipe geometry. The flow velocity is denoted by arrows, showing the typical "plug" profile arising from the increased haematocrit towards the center of the vessel.*

These gradients have significant effect on the trafficking of the cells and therefore they influence the distribution of the cells and the overall rheology. A well-known implication is the emergent phenomenon of platelet margination. RBCs move away from the vessel walls primarily due to the effect of wall-induced lift and they create a haematocrit distribution that increases towards the center of the channel. This in turn pushes platelets out of the center of the vessel towards the wall through shear-induced cell-cell collisions. Every time a platelet bumps into an RBC it is displaces a little (see Fig. 6). This displacement mechanism favours the direction towards the wall, since the RBC density is lower there.

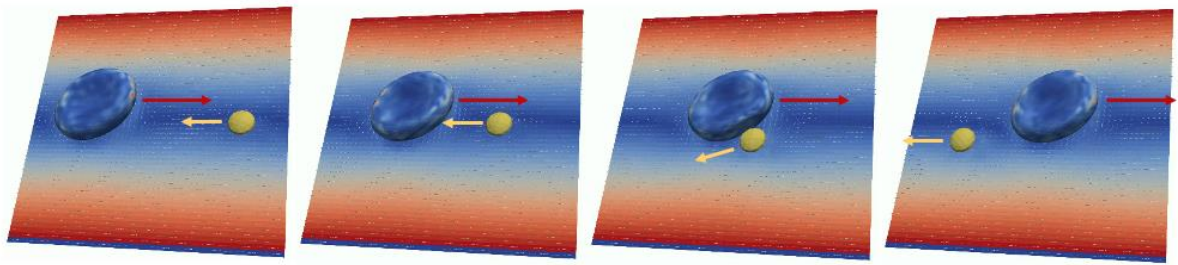

*Figure 6 Displacement of a platelet during a collision with a red blood cell. Time progresses from left to right. The platelet is displaced by approx. 1 µm during the collision.*

After a long sequence of collisions, platelets tend to get captured in the CFL at the wall. In an evolutionary sense this is a very important mechanism, since it drives PLTs next to the vessel wall, where they need to act in case of an injury to stop the bleeding. When this mechanism is impaired, it leads directly to bleeding disorders as it was demonstrated recently using HemoCell in combination with a series of bleeding experiments [35].

### 3.2 Cellular flow in microfluidic devices

Biochemical processes of blood are often investigated in microfluidic devices (also called "bleeding chips"). In recent years the role of the biomechanical environment is gaining focus as an inseparable component acting along biochemistry. See for instance the mechanism of shear induced thrombus formation [36], that produces the majority of stroke and heart infarct cases. While the role of the mechanical forces are becoming obvious through mechanosensing components, such as the von Willebrand factor (VWF) [37], their quantitative description is lacking. The primary reason for this is the limitation in current measurement technologies to capture detailed stresses in fluid and cells. Numerical methods can yield the missing complimentary information by replicating the microfluidic flow environment of the measurements in high detail. HemoCell was used to investigate the starting point of forming high-shear rate thrombi to explore what are the necessary conditions that allow such a formation [38]. Multiple flow conditions were examined in multiple micro-stenosis geometries. Cross-matching the cellular simulations with the experimental results revealed that at the initial location of the thrombus multiple conditions need to be present: 1) Availability of platelets (the high availability is facilitated by margination). 2) Suitable "collision-free" volume, that can be provided for example behind a stenosis by the mechanism of cell-free layer formation. 3.) (In case of a shear induced thrombus) high shear rate values that can uncoil and activate VWF. Apart from such a suitable mechanical environment, the components of Virchow's triad, including the thrombogenic surface, are still necessary for the development of the complete thrombus.

The cellular nature of blood significantly influences the mechanical responses, for instance the emerging stresses. Current empirical blood models that use a continuum approximation cannot include the effects of non-homogeneous cell distributions. When investigating the biomechanics of the vessel wall (for wall remodelling, inflammation, endothelial layer disruption, etc.) wall shear stress is one of the key quantities of interest. Continuum approaches cannot resolve the CFL, and as a consequence they underestimate the wall shear rate, and at the same time they tend to significantly over-estimate wall stresses. The reason for this misprediction is that the CFL (forming next to the wall) is a pure plasma layer, that has a viscosity three times lower than the average whole blood viscosity. This natural lubrication layer influences the local flow dynamics significantly, that cellular models can capture successfully [39]. It is also worth noting that most microfluidic experiments are characterised based on wall shear rate or stress calculated using continuum theories (such as Hagen-Poiseuille or Poiseuille flow). These calculations likely have the same shortcoming since they neglect the CFL.

### 3.3 Flow in a curved micro-vessel section

Traditionally, cellular blood flow mechanics during platelet adhesion and aggregation are studied in straight channels, sometimes including local geometric variations (e.g., stenoses)

[39], while the influence of more complex vessel geometries on blood flow characteristics remains understudied. The effects arising from more complicated geometry are discussed here focusing on initial platelet adhesion and aggregation, via a combination of *in silico* (i.e., HemoCell) and *in vitro* approach. The flow behaviour in regard to shear rate and rate of elongation as well as cellular distributions are investigated in the simulations of a curved channel and compared to the results of complementary microfluidic aggregation experiments in a similarly shaped flow chamber. The simulations reveal the occurrence of high elongational flow at the inner arc of the curvature which corresponds to the site of increased platelet aggregation in the experimental results.

The simulations are performed in a U-shaped square duct geometry with a 25*25 µm$^2$ cross-section, and the curvature has an inner diameter of 25 µm as well. The domain is initialised with RBCs to result in a discharge haematocrit of 30%. In order to ensure a constant inflow from a straight vessel section a periodic pre-inlet is utilised (see Fig. **7** A). The blood flow is simulated for two seconds at two different flow velocities which are indicated by their initial wall shear rates: 300 s$^{-1}$ and 1600 s$^{-1}$. For the microfluidic aggregation experiments a similarly shaped U-chamber is designed at a larger scale. The chamber is coated with collagen and perfused for 4 minutes with hirudinated human whole blood at the same two flow velocities as used in the simulations. More detailed setup methodology can be found in [40].

In order to capture effects of the curvature, cell distributions, width of the CFL and cross-sectional flow profiles are quantified in three sections: close to the inlet, at the center of the curvature and close to the outlet. While differences in CFL width remain insignificant and only a slight shift of RBC concentration towards the inner wall of the curvature is observed, the flow profiles display significant differences when comparing the curvature to the inlet region. Increased shear rate gradients which cause sites of high elongational flow are observed at the inner arc of the curvature (see Fig. **7** B and C). When comparing the two flow velocities, we observe qualitative similarity with different magnitudes in each case.

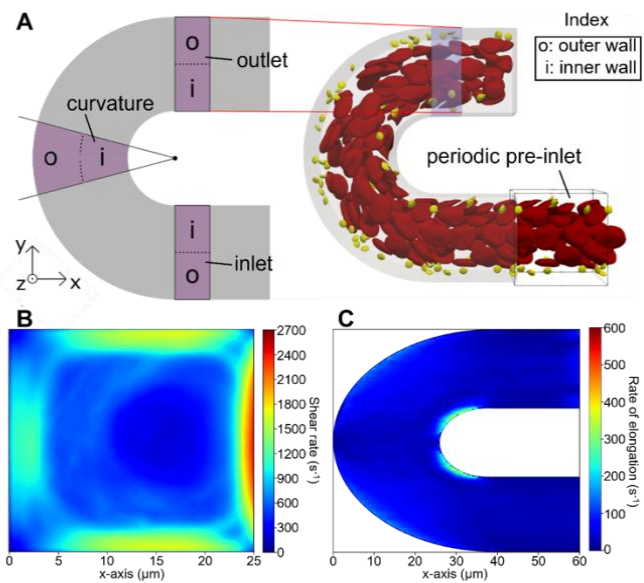

*Figure 7 Curved channel simulations. A) Setup of curved channel domain with flow in negative x-direction from periodic pre-inlet. The regions of interest are highlighted with the corresponding inner and outer wall division. B) Cross-sectional shear rate profile in curvature section and C) top view of elongational flow magnitude across channel at initial wall shear rate of 1600 s-1.*

To allow for a comparison of the simulations to the experiments, the aggregate sizes accumulated on the microfluidic chamber after perfusion are quantified in the same respective regions, denoted inlet, curvature and outlet. The results show different observations for the two flow velocity cases. While the low flow velocity chamber displays no significant difference in aggregate size between the different regions, the high flow velocity chamber depicts elevated aggregate formation at the inner arc of the curvature.

The increased aggregate area of the microfluidic experiments is located at the sites of high shear and elongational flows, observed in the simulations. As the blood used in the experiments is treated with the anticoagulant hirudin, only the initial steps of platelet adhesion and aggregation are observed, which highly depend on the plasma suspended molecule von Willebrand factor as a mediator [41]. While shear-induced platelet aggregation occurs at shear rates much higher than observed in the simulations [42], elongational flows are found to enable the von Willebrand factor-mediated adhesion process at comparatively low shear flow [43]. Since the peak rate of elongation in the simulations reaches this critical range only in the high flow velocity case, it is hypothesized to be responsible for the elevated aggregate formation, which is also only observed at the higher flow velocity in the experiments. In summary, the results highlight the role of elongational flows as well as the importance of vessel geometry in initial platelet adhesion and aggregation.

### 3.4 Flow of diabetic blood in vessels

Several diseases can alter the deformability of cells significantly, and therefore impact the overall rheology of whole blood. The deformability of a red blood cell can be impeded, and stiffened, by various pathologies such as sickle cell disease [44], diabetes [45], malaria [46], and Parkinson's disease [47] for example. In diabetec flow RBCs are stiffer, less capapble of deformations which have a significant effect on bulk viscosity as well as on platelet margination. A novel stiffened red blood cell model was applied to study the effects of changing red blood cell deformability. This model was developed and validated in HemoCell by matching the deformation indices from ektacytometry measurements *in vitro* with the deformation indices computed from single-cell shearing numerical experiments *in silico* [27]. Stiffening of RBCs membranes was induced *in vitro* by incubating RBCs with tert-butyl hydroperoxide (TBHP), which induces oxidative stress on the cell membrane. TBHP can be used as a general model for cell membrane perturbations resulting from oxidative disorders

[48]. The stiffened numerical RBC model was achieved by scaling the mechanical parameters, in particular the link force coefficient and the internal viscosity ratio of the original validated RBC model [26], [49].

The behaviour of stiffened RBCs in bulk rheology, was studied by simulating whole blood flowing through a periodic pipe of radius R = 50 μm with a tank haematocrit of 30% driven by a body force resulting in a wall shear rate of 1000 s$^{-1}$. The fraction of stiff/healthy RBCs (0/100, 30/70, 50/50, 70/30, and 100/0) was varied in each simulation, maintaining a total haematocrit of 30%. The main observation from this study is that the RBC free layer decreases because of an increase of stiffened RBC fraction in flowing blood, shown in the left panel of Fig. 8. The more rigid RBCs experience decreased lift force leading to reduction in the CFL size. Furthermore, a decrease of platelet localization at the vessel wall as the fraction of stiffened RBCs increase is also observed, right panel of Fig. 8.

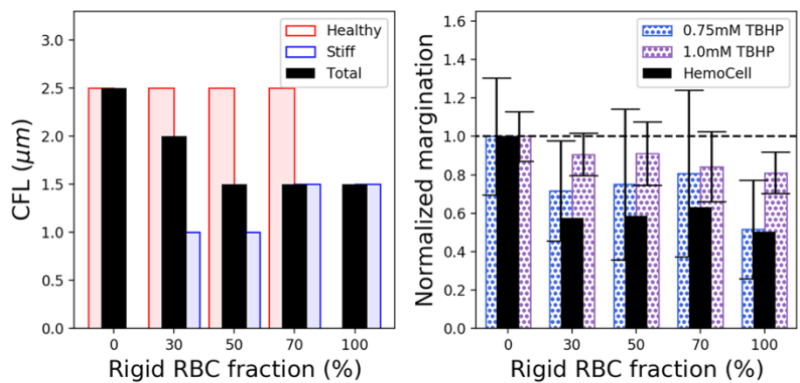

*Figure 8* Red blood cell-free layer (left panel) and platelet margination (right panel) as a function of rigid RBC fractions. The computed CFL from HemoCell is shown in black, with each of the RBC components shown in red for healthy and blue for 1.0mM TBHP. Platelet concentration at the wall is computed in the volume 4 μm from the wall normalized to the concentration of the 100% healthy RBC case (HemoCell:black and in vitro results for 0.75mM TBHP:blue and 1.0mM TBHP:purple).

Platelet margination is likely altered on the one hand by the decrease in size of the CFL as there is limited volume next to the wall for platelets to be trapped in, and on the other hand by a reduced shear induced dispersion that would drive the PLTs towards the CFL. Note that these changes might have an influential effect on the haemostatic processes, where the high availability of PLTs is one of the necessary conditions for physiologic operation.

The work presented in this section proposes a general model for altered RBC deformability as a result of disease, both experimental and computational, and offers evidence of the detrimental effect rigid RBCs have on physiological blood flow and the ability of platelets to localize to the vessel wall. The diseased, stiffened, red blood cell model from this section has further been applied to investigate whole blood flow through a patient specific segmented retinal microaneurysm [50].